\documentclass[doublecol]{epl2}
\usepackage{latexsym}
\usepackage{graphicx}
\usepackage{color}
\usepackage{amssymb}
\usepackage{amsmath}
\usepackage{amsfonts}
\newcommand{\pa}[1]{\left(#1\right)}

\newcommand{\newc}{\newcommand}
\newc{\beq}{\begin{equation}}
\newc{\eeq}{\end{equation}}
\newc{\kt}{\rangle}
\newc{\br}{\langle}
\newc{\beqa}{\begin{eqnarray}}
\newc{\eeqa}{\end{eqnarray}}
\newc{\pr}{\prime}
\newc{\longra}{\longrightarrow}
\newc{\ot}{\otimes}
\newc{\rarrow}{\rightarrow}
\newc{\h}{\hat}
\newc{\bom}{\boldmath}
\newc{\btd}{\bigtriangledown}
\newc{\al}{\alpha}
\newc{\be}{\beta}
\newc{\ld}{\lambda}
\newc{\sg}{\sigma}
\newc{\p}{\psi}
\newc{\eps}{\epsilon}
\newc{\om}{\omega}
\newc{\mb}{\mbox}
\newc{\tm}{\times}
\newc{\hu}{\hat{u}}
\newc{\hv}{\hat{v}}
\newc{\hk}{\hat{K}}
\newc{\ra}{\rightarrow}
\newc{\non}{\nonumber}
\newc{\ul}{\underline}
\newc{\hs}{\hspace}
\newc{\longla}{\longleftarrow}
\newc{\ts}{\textstyle}
\newc{\f}{\frac}
\newc{\df}{\dfrac}
\newc{\ovl}{\overline}
\newc{\bc}{\begin{center}}
\newc{\ec}{\end{center}}
\newc{\dg}{\dagger}
\newc{\prh}{\mbox{PR}_H}
\newc{\prq}{\mbox{PR}_q}
\newc{\tr}{\mbox{tr}}
\newc{\pd}{\partial}
\newc{\qv}{\vec{q}}
\newc{\pv}{\vec{p}}
\newc{\dqv}{\delta\vec{q}}
\newc{\dpv}{\delta\vec{p}}
\newc{\mbq}{\mathbf{q}}
\newc{\mbqp}{\mathbf{q'}}
\newc{\mbpp}{\mathbf{p'}}
\newc{\mbp}{\mathbf{p}}
\newc{\mbn}{\mathbf{\nabla}}
\newc{\dmbq}{\delta \mbq}
\newc{\dmbp}{\delta \mbp}
\newc{\T}{\mathsf{T}}
\newc{\J}{\mathsf{J}}
\newc{\sfL}{\mathsf{L}}
\newc{\C}{\mathsf{C}}
\newc{\B}{\mathsf{M}}
\newc{\V}{\mathsf{V}}

\title{Record statistics in random vectors and quantum chaos}

\author{Shashi C. L. Srivastava\inst{1} \and Arul Lakshminarayan\inst{2} \and Sudhir R. Jain\inst{3}}
\shortauthor{S. C. L. Srivastava \etal}
\institute{
  \inst{1} RIB Group, Variable Energy Cyclotron Centre, 1/AF Bidhan nagar, Kolkata 700 064, India\\
  \inst{2} Department of Physics, Indian Institute of Technology Madras, Chennai 600036, India\\
  \inst{3} Nuclear Physics Division, Bhabha Atomic Research Centre, Mumbai 400 085, India
}

\date{\today}

\abstract{
The record statistics of quantum standard map is shown to capture the classical transition to chaos. It is shown  that in the mixed phase space regime the number of intensity records
is a power law in the dimensionality of the state as opposed to the logarithmic growth for the random states. The exponent of this power-law is close to $0.5$ at the critical value of the chaos parameter, $K \simeq 0.98$ of the standard map, an exponent that is also obtained for random walks. These findings are based on the record statistics of complex, normalized random states for which we have shown that the probability of a record intensity is a Bernoulli process.}

\begin{document}
\maketitle

\section{Introduction}
The success of the many books and websites on record breaking events is a testimony of their enduring fascination. In an indexed data set, or discrete time series, a ``record" is an event that is larger (upper record) or smaller (lower record)  than all preceding ones \cite{Nagaraj}. In the following upper records are exclusively studied and are simply referred to as records. Questions about how the records increase with time, or the number of records set are of a natural interest
in a variety of contexts from sports to rainfall \cite{Gembris02, Vogel01,Redner06,krug10,sanjeev}, and a  mathematical theory of records for independent identically distributed (i.i.d.) random variables has been developed \cite{Renyi,Glick78,Nev}.
If $\{x_t, t=1, \cdots N\}$ is a finite time series, the first element, $R_1$, of the corresponding records series is $x_1$ itself and at subsequent times $t$ it will be $R_t=\mbox{max}(x_t,R_{t-1})$. As $x_t$ is a random variable, so is $R_t$ and properties of this random variable are of interest.

The study of the statistics of  extreme events plays a very important role in a variety of  contexts from hydrology to climate change \cite{Gumbel}. Apart from numerous statistical problems to which it naturally applies, the theory of extremes has also been used in the context of deterministic dynamical systems \cite{Nicolis06}. One of the motivations of studying extreme values in quantum states is to understand whether deviations arising for dynamical systems  originate from the localization of wavefunctions. This, in turn, is connected to the important issue of quantum ergodicity. The Tracy-Widom distribution \cite{Tracy96} of the extreme  eigenvalues of random matrices is also relevant to finding the fraction of entangled states \cite{Uday12}. In \cite{Arul08}, the statistics of maximum and minimum intensities in complex random states, and, in the quantum standard map have been studied. However the study of records in random states and quantum dynamical systems has not yet been explored, and is the
intention of this Letter.

The statistical nature of quantum chaotic states have so far been explored mainly with the random matrix theory (RMT) \cite{Mehta04,Brody81}, and while indeed this has been successful, there remain many questions, especially for cases in which the classical dynamics is not fully chaotic, and also for many-body systems with few-body interactions. Deviations from universal RMT predictions may have interesting dynamical origins.
Records is a relatively new statistical tool that is interesting because some of the features like number of records is ``distribution free'' for i.i.d. processes, that is are independent of the underlying probability distribution for the variables, and therefore any deviations in this is a reflection of correlations.  Large (or small) amplitudes is interesting from the point of view of localization, and therefore the study of extremes and records for wavefunctions is of interest. Indeed beyond quantum chaos, the theory of records can be of interest in the study of localization in general, including Anderson localization.

Due to the complexity of the eigenfunctions of chaotic quantum systems, it is expected that the records in the intensity  would be distributed as those for random vectors. It may be noted that while the present work considers the intensities of the quantum states, this is interesting in itself as the statistical properties of the probabilities are associated with complete von Neumann position measurements. At the same time, the phases are not given a complete go by, as the momentum representation is simultaneously studied.  Indeed it is well-known that the momentum and position intensities together encapsulate essential quantum features such as the uncertainty principle \cite{Bialynicki}.

Here general results on record statistics for a ``time-series" employing the intensities of wavefunctions is obtained. These include results on interesting new quantities such as the number of records in a state, and the distribution of the {\it position} of the maximum.  In the following, record statistics for normalized random states are studied and subsequently the record statistics in the eigenstates of the quantum standard map - a paradigm of classical and quantum chaos - are shown to reflect the nature of classical phase space, whether regular, mixed or chaotic. Significantly, not only is the transition to complete chaos reflected in such statistics, but the breaking of the last golden mean KAM tori is seen as a critical event in a way that will be elaborated.

\section{Record statistics for $\delta$-correlated variables}
Let $|\psi\kt$ be a normalized state vector, with components $z_n = \br n|\psi\kt$ in a complete orthonormal  basis $|n \kt$. The ``time-series"  considered is simply $x_n=|\br n |\psi\kt |^2$, $n=1, \ldots, N$, where $N$ is the dimension of the Hilbert space. Thus the ``time" in which the records are observed is not the real time, but could be some observable such as position or momentum. The question of records is then about statistics of large intensities, their positions, occurring along the index.
Define the probability density for the record variable to be $R$, at time $t$ as $P(R,t)$. The average record $\langle R_t \rangle $ is given by the first moment.
Let $P(x_1,\ldots, x_N)$ be the j.p.d.f. of $N$ random variables. The probability that the record at time $t$, $R_t$, is less than $R$ is given by
\begin{equation}
Q(R,t)= \int_{0}^{R} dx_1...dx_t P_t(x_1, ...x_t)
\end{equation}
where $P_t(x_1,\ldots,x_t)= \int P(x_1, \ldots,x_N) dx_{t+1}...dx_{N}$ is the marginal j.p.d.f. of the first $t$ random variables.  It follows that
$P(R,t) =dQ(R,t)/dR$.

Components of normalized complex random vectors $z_i=\br n |\psi\kt$, have the j.p.d.f.:
$P(z_1,z_2, \ldots, z_N) = (\Gamma(N)/\pi^N) \delta \left(\sum_{j=1}^N |z_j|^2-1\right).$
This is also the distribution of the components of the eigenvectors of the GUE or CUE (Gaussian or Circular unitary ensembles) random matrices. It is the invariant uniform distribution with respect to the Haar measure under arbitrary unitary transformations on the $(2N-1)$-dimensional sphere. The correlation induced by normalization becomes weak for large $N$. The intensities $x_i=|z_i|^2$ being the random variables, it is useful to define the j.p.d.f. as
\begin{equation}\label{eq:jpdfgue}
P(x_1,\dots, x_n;u) = \Gamma(N)\delta\pa{\sum_{i=1}^N x_i -u},
\end{equation}
where $u$ is an auxiliary quantity - the actual j.p.d.f. corresponding to $u=1$.
Defining
\beq
Q(R,t;u) = \int_{0}^{R} \prod_{k=1}^{t}\upd x_k \int_0^{\infty} P(x_1,\dots, x_N;u) \upd x_{t+1}...\upd x_{N},
\eeq
leads to
\beq
\int_0^{\infty} e^{-su} Q(R,t;u) du =\frac{\Gamma(N)}{s^{N}} \sum_{m=0}^t (-1)^m \binom{t}{m} e^{-s mR} .
\eeq
Using the convolution theorem, and then setting $u=1$ in $Q(R,t;u)$ gives
\begin{equation}\label{eq:qrt}
Q(R,t) = \sum_{m=0}^t (-1)^m \binom{t}{m} (1 - mR)^{N-1} \Theta (1-mR),
\end{equation}
Hence $P(R,t)=
\sum_{m=1}^t (-1)^{m+1} \binom{t}{m} m (N-1)(1 - mR)^{N-2} \Theta (1-mR)$,
the probability density that the record is $R$ at time $t$. Note that the final record is also the global maximum of the sequence, hence the final-record statistics provides another way to calculate the distribution of the maximum and it agrees with results obtained in \cite{Arul08} for random GUE vectors. A word of caution is however in order, only the last record and the global maximum coincide with each other. It should be appreciated that the second record and the second largest member the sequence are {\it not} the same. Hence, record distribution is a different measure and should not be seen as a  generalization of the work in \cite{Arul08}.

The $\delta$-correlated process studied here is a special case of the ``product measure state'' whose j.p.d.f. is $P(x_1,x_2,...,x_N)= C_N \, \prod_{i=1}^N f(x_i) \delta(\sum_{i=1}^N x_i-M)$, where $f(x_i)$ is some function,  $M$ a mass parameter and $C_N$ the overall normalization. These arise as the steady states in various dynamic models of non-equilibrium statistical physics such as the zero-range process for which record statistics is still unknown. However, it has been shown that for $f(x_i)\simeq A x^{-\gamma}$, the one variable marginal distribution undergoes a phase transition by increasing the density $\rho~ (=M/N)$ for $\gamma > 2$ \cite{MEZ}. This phase transition from fluid phase to anomalous or normal condensate depending on the parameters $\gamma$ and $\rho$, reflects in extreme mass distribution too. The extreme value distribution changes from Gumbel, via Fr\'echet to a non-standard extreme value distribution as $\rho$ changes from sub-critical via critical to super-critical values\cite{evans}.  Thus the present study of records in a special case has wider motivations in such contexts as well.

Returning to the problem at hand, it can be easily seen that  the $t^{th}$ record is Gumbel distributed for large $N$, as for large $N$ and $t\gg 1$, Eq.~(\ref{eq:qrt}) becomes
\beq
Q(R,t) \approx (1-\exp(-NR))^t \approx \exp\left(-t \exp(-NR) \right).
\eeq
Since the Gumbel distribution is of the form $\exp[-\exp [-(x-a_N)/b_N]]$ where $a_N$ and $b_N$
are the shift and scaling, it follows that for the records statistics the relevant parameters are
$a_N=\log(t)/N$ and $b_N=1/N$. The above form also appears in the limit when the correlations are ignored.

The average record as a function of time is $\langle R(t)\rangle=1-\int_0^1 Q(R,t)\, dR$ which is equal to
\beq \label{eq:avRgue_exact}
\frac{1}{N}\sum_{m=1}^t (-1)^{m+1}  \frac{1}{m} \binom{t}{m} = \f{H_t}{N}=\frac{1}{N}\sum_{k=1}^t \f{1}{k},
\eeq
where $H_t$ is a Harmonic number as defined above.
Known asymptotics of the Harmonic numbers implies that
\begin{equation}
\label{randomavgrecord}
\br R(t) \kt  = \f{1}{N}\left( \gamma+ \ln(t) + \frac{1}{2t} - \sum_{k=1}^{\infty} \f{B_{2k}}{2k\,  t^{2k}} \right),
\end{equation}
where $B_{2k}$ are Bernoulli numbers, and $\gamma$ is the Euler-Mascheroni constant. Again, this presents a generalization of the average maximum intensity found in \cite{Arul08} which corresponds to $t=N$.

Other interesting quantities include the average number of records, the distribution of the lifetimes of the records, the probability that the last record (which is also the maximum) survives for a given time and so on. A remarkable well-known fact from the theory of records is that for i.i.d. variables these quantities are distribution-free, that is independent of the particular underlying distribution $p(x)$ \cite{Nagaraj}. For example the average number of records $\br N_R\kt = H_N\sim \log(N)+ \gamma$ is indeed very small compared to the length $N$ of the data set; typically records are rare events. These follow from a classic result \cite{Renyi,Nagaraj} that the probability of a record occurring at position $j$ is $1/j$, independent of the past and future position of the records. In other words the probability of the position of the records is a Bernoulli process, $\mbox{Ber}(1/j)$.

  As the rank order of elements distributed in an i.i.d. fashion does not change due to a normalization induced delta function correlation, not surprisingly, the average number of records for random vectors (and also in case of product measure state) is, and can be proven to be, {\it same} as the $\br N_R\kt$ stated above. It is interesting to note that while the record values do depend, for finite $N$, on the global constraint, the number of records do not. Being distribution-free, the  statistics of records is a direct measure of correlations. There are few analytical results for correlated variables, one notable exception being a random walk where it has been shown that the number of records grows much faster as $\sqrt{N}$ while the probability that a record occurs at $j$ decays as $1/\sqrt{j}$ \cite{Majumdar08} rather than the i.i.d. and random states results of $\log(N)$ and $1/j$ respectively. In the case of the random walk  the number of records is not a self-averaging quantity, the
standard deviation being of the order of the mean.

\section{Records for eigenvectors of quantum standard map}
We turn from random vectors to eigenvectors of the quantum standard map. Classically, the standard map on the unit torus is an area-preserving map: $(q',p')= (q+p,\,p-(K/2\pi) \sin(2 \pi q'))$ where as the parameter $K$ changes from zero to unity, the system undergoes a transition from integrable (free rotor) to a mixed system. At about $K\approx 1$, the last rotational  KAM torus breaks allowing global diffusion in the momentum space \cite{ll}. If the standard map is unfolded to a cylinder, it displays normal diffusion in momentum for large enough $K$. When $K \gg 5$, the classical map is essentially fully chaotic. For such parameters this leads to quantum eigenstates which follow the CUE/GUE or COE/GOE results depending on the value of the phases $\alpha$ and $\beta$, which control parity symmetry and time-reversal symmetry respectively. One may think of $\beta$ as originating from a threading magnetic flux line, while $\alpha$ originates from the potential itself.   If $\beta \ne 0$ and $\alpha \ne    0,1/2$ we can expect that both the time-reversal symmetry and parity symmetry are broken and the typical eigenstates would be like complex random states. For the record statistics, we calculate the statistical properties of eigenvectors of the
Floquet operator  \cite{izra} in the position basis of a kicked pendulum on a torus: $U_{nn'}=$
\begin{equation}\label{eq:uniop}
\begin{split}
&\frac{1}{N} \sum_{m=0}^{N-1} \exp\left[ -i \pi \f{(m+\beta)^2}{N} + 2 \pi i \f{(m+\beta)}{N}(n-n') \right ]\\ &\times \exp \left[ i \f{K N }{2 \pi}\cos \frac{2\pi (n+\alpha)}{N} \right].
\end{split}
\end{equation}

The dimensionality of the Hilbert space $N$ is the inverse (scaled) Planck constant. Thus the ``data'' in this case are the various eigenfunctions and especially their intensities.  Clearly this can, in general, depend on the space in which the eigenfunctions are represented. Thus for small values of $K$ we expect there to be many localized states in the momentum space while being nearly uniformly distributed in the position, and this will reflect in any studies of records or extremes. However for large $K$, position or momentum basis will be equivalent up to fluctuations. For all the numerical calculation, unless otherwise stated, $N=400$, $\alpha=0.25$ and $\beta = 0.25$ are taken.
\begin{figure}[htb]
\onefigure[scale=0.5]{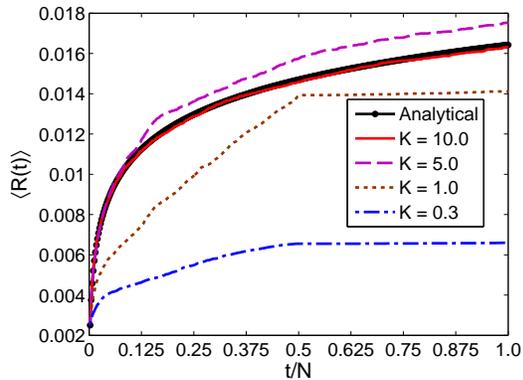}
\caption{The average record $\langle R(t)\rangle$, from the ensemble of eigenstates of the quantum standard map in the position representation. The parameters used are $N=400$ and $K=10$ (highly chaotic), $K=5$ (mostly chaotic), $K=1$ (mixed phase space), and $K=0.3$ (mostly regular). The analytical curve refers to the random state result in Eq.~(\ref{randomavgrecord}). In all cases $\alpha=\beta=0.25$.}
 \label{fig:rt_tbyN}
\end{figure}

\begin{figure}[htb]
\onefigure[scale=0.5]{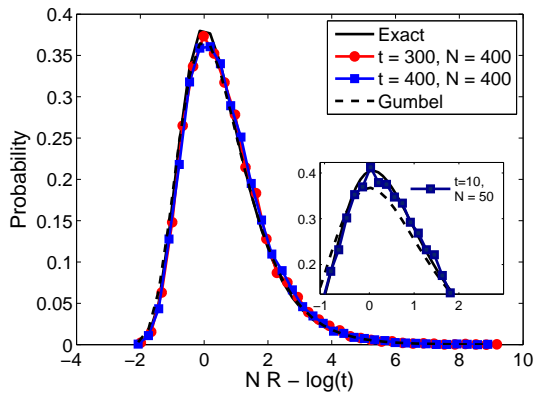}
\caption{ The distribution of the records when the index is $t$ for eigenfunctions of the quantum standard map with $K=10$. After rescaling and a shift, the distributions are of the Gumbel type, except for small $N$ (see inset) where deviations are seen and the exact formula for $P(R,t)$ is to be used. }
 \label{fig:Gumbel}
\end{figure}

The average record as a function of the index normalized by the dimension of the Hilbert space (which acts as ``time'' for  these vectors)  is plotted in Fig.~\ref{fig:rt_tbyN} for various values of $K$.  While this agrees well in the chaotic region with the result on random states, there are significant deviations in the mixed phase space regime of  $K \lesssim 5$; eventually, these disappear for $K = 10$. For  $K$ $(=0.3)$, in the position space,  most of the records are set up by $t/N=0.5$, originating in the very weakly broken parity symmetry. The momentum space
average records are somewhat similar but mostly, they lie above the result for random states and are not affected much by the broken parity symmetry due to their localization. Indeed, the record for eigenfunctions of the quantum standard map in the classically chaotic regime is found to be Gumbel-distributed, see Fig.~\ref{fig:Gumbel}. Here, we have also plotted  the distribution for the ``record" when $t=N$, which refers to the maximum intensity, for which the explanation is found in \cite{Arul08}. For small $N$, deviations from the Gumbel distribution are evident. These are consistent with the exact result, $P(R,t)$ derived above (shown in the inset of Fig.~\ref{fig:Gumbel}).

\begin{figure}[htb]
\onefigure[scale=0.5]{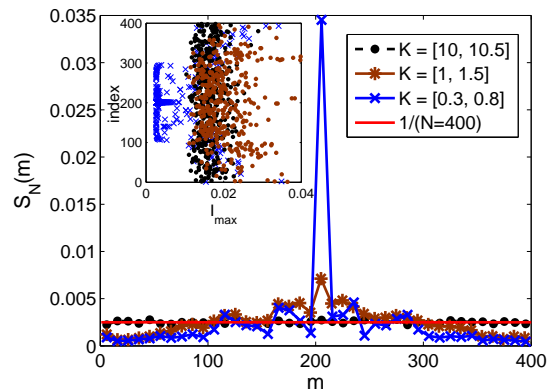}
\caption{The distribution of the position of the last record set, which is also the maximum, for eigenfunctions of the standard map with $N=400$ and for various values of $K$. The inset shows the maximum intensity $I_{\mbox{max}}$ of individual states and where it occurs.}
 \label{fig:SN_m}
\end{figure}

The distribution of the position of the maximum in the position representation is shown in Fig.~\ref{fig:SN_m}, where one can see a transition to the uniform distribution along with the transition to classical chaos. The sharp peak at the center for small $K$ (here $K=0.3$) is interesting and deserves further comment. For small $K$, when there are many narrow classical resonances, a significant fraction of eigenfunctions that are localized on separatrices possess maximum intensity at or very close to $q=1/2$. For instance, for $N=400$ when $K=0.1$ and $0.3$, about $75\%$ and $50\%$ of states are peaked at $q=1/2$.  As we increase $K$, more islands start appearing with turning points away from $q=1/2$, and eigenstates localize on their interior, thereby the maximum intensity shifts away from $q=1/2$. This provides a qualitative mechanism leading to  uniform distribution for maximum intensity for large $K$. The transition in classical behavior from integrable ($K=0$), to mixed phase space where islands of
stability coexist within the stochastic sea (intermediate $K$), to fully-developed chaos is well captured by the quantity $S_N(m)$. In the inset of Fig.~\ref{fig:SN_m} is shown the maximum intensity of individual states (again $N=400$) vs their actual position. For $K = 0.3$, there are a large number of states which have maximum intensity at q = 1/2 (index = 200). This manifests in the form of the observed sharp peak. These get gradually destroyed on increasing $K$; indeed such measures seem useful to pursue further in classifying states in the mixed phase-space regime.

\begin{figure}[htb]
\onefigure[scale=0.5]{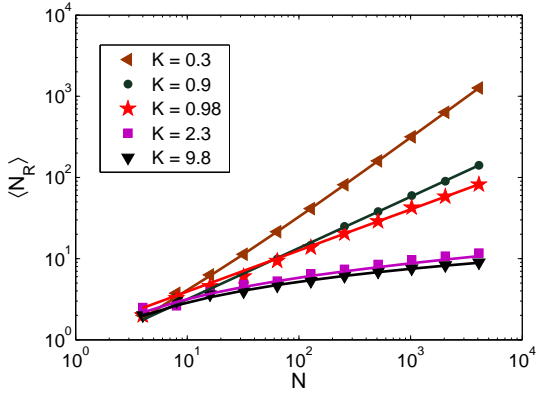}
\caption{ The average number of records in the eigenfunction of the quantum standard map of length $N$ vs N is plotted for various $K$ values. Solid line (except for $K=9.8$) are fitted expression $a\log(N)+bN^{\delta}$, for $K=9.8$ the solid line is $\log(N)+\gamma$. $a$, $b$, $\delta$ for different $K$ values are,(I) $K=0.3$, $a=0.45\pm 0.15$, $b=0.306\pm 0.007$, $\delta=1.001\pm0.003$,(II) $K=0.9$, $a\approx 0$, $b=0.73\pm0.06$, $\delta=0.63\pm 0.01$, (III) $K=0.98$, $a\approx 0$, $b=1.2\pm0.2$, $\delta=0.50\pm0.01$ and (IV) $K=2.3$, $a=0.9\pm0.3$, $b=0.7\pm 0.6$, $\delta=0.21\pm0.04$.}
 \label{fig:NRvsN}
\end{figure}

It has been shown above that for random $N$-dimensional vectors  there are on the average $\sim \log N+\gamma$ intensity records.  While we can expect to see this for the standard map in the chaotic
regime, the mixed and near-integrable regimes show a marked departure. The correlations lead to results that are similar to those for the random walks, with a power-law scaling in $N$. The ratio of standard
deviation of the number of records to the average is found to be of order unity, indicating non-stationarity of the distribution. For very small $K$ the number of records is simply  $\mathcal{O}(N)$, as the eigenfunctions
are describable by smooth functions. In the mixed-regime, and, when $K < 1$, a power-law along with a logarithmic dependance is clearly indicated. Intriguingly, {\it it is almost a pure power-law with exponent $0.5$ at the critical value $K = 0.98$}. For moderate values such as $K=2.3$ a clear separation of the phase space into chaotic and regular regions and a subsequent separation of the quantum spectra seem to make the numerical fits unstable. Further, for very large $K$ (say 9.8), the result from the random vector case applies (Fig. \ref{fig:NRvsN}).
This reflects the changing nature of the wavefunction intensities, and indicates that at criticality it is close to some kind of $1/f$ noise. Eigenvalue fluctuations in quantum chaotic systems are known to have $1/f^{\gamma}$ noise \cite{Santhanam05}, and further work on the spectral properties of intensities is therefore called for.

\begin{figure}[htb]
\onefigure[scale=0.5]{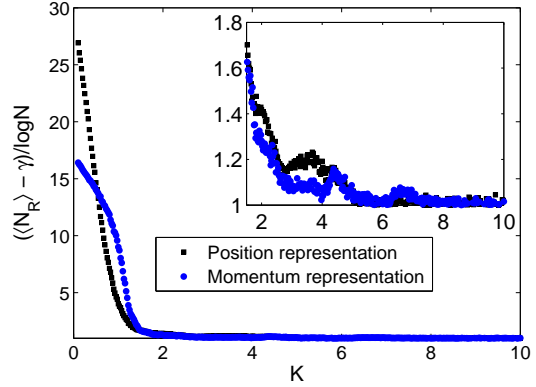}
\caption{ An appropriately scaled and shifted average number of records $\langle N_R\rangle$ {\it vs} $K$ for the eigenfunctions of the quantum standard map with $N=400$.}
 \label{fig:rN_K}
\end{figure}

The average number of records as a function of $K$ is presented in Fig.~\ref{fig:rN_K}, where the effects of the scaling in the mixed regime manifest themselves. It is remarkable that while record numbers in both, the momentum
and the position representation,  approximately converge after $K \approx 2$, it is only for  $K > 5$ when there may be tiny islands (if at all) that it returns to the result of the random vectors. These observations establish  the connection between classical dynamics in all the regimes with the record statistics for the eigenvectors of the quantum standard map. Since the standard map is a paradigmatic model, we believe this connection to hold generally.

\section{Conclusions}The probability that a record appears in the intensities of correlated random vectors derived results on record of intensities of correlated random vectors at an index $j$ is a Bernoulli process, the same as for i.i.d. variables. The quantum standard map presents itself as a generic instance where eigenfunctions become increasingly complex with the system parameter, $K$. Remarkably, a signature of breaking of the last torus at $K \simeq 0.98$ is found to be in the exponent one-half in the power-law obeyed by the number of records.  It is very likely that the conclusions drawn on the quantum standard map would hold for other quasi-integrable and mixed systems.

Various extensions are possible - for higher dimensions, the theory of records  has a multivariate extension \cite{Nagaraj}; the role of simultaneous breakdown of parity and time-reversal on records of intensities would be worth studying too.

\end{document}